\documentclass[%
superscriptaddress,
 amsmath,amssymb,
 aps,
pra,
]{revtex4-1}

\usepackage{amsthm}
\usepackage{graphicx}
\usepackage{dcolumn}
\usepackage{bm}
\usepackage{multirow} 
\usepackage{enumerate}
\usepackage{algorithm}
\usepackage{algorithmic}
\usepackage{tabularx}
\usepackage{hyperref}
\usepackage[mathlines]{lineno}%
\usepackage[roman]{complexity}

\newcommand{\abs}[1]{\left\lvert#1\right\rvert}
\newcommand{\norm}[1]{\left\lVert#1\right\rVert}
\newcommand{\ket}[1]{\left|#1 \right\rangle}
\newcommand{\bra}[1]{\left\langle #1 \right|}
\newcommand{\braket}[2]{\left\langle #1| #2 \right\rangle}
\newcommand{\ketbra}[2]{\left| #1 \rangle \langle #2 \right|}

\theoremstyle{remark}

\begin{document}

\makeatletter
\newcommand{\rmnum}[1]{\romannumeral #1}
\newcommand{\Rmnum}[1]{\expandafter\@slowromancap\romannumeral #1@}
\newcommand{\sgn}{\text{sgn}}
\newcommand{\Tr}{\textmd{Tr}}

\makeatother

\preprint{APS/123-QED}
\title{Quantum data compression by principal component analysis}

\author{Chao-Hua Yu}
\email{quantum.ych@gmail.com}
\affiliation{State Key Laboratory of Networking and Switching Technology, Beijing University of Posts and Telecommunications, Beijing 100876, China}
\affiliation{State Key Laboratory of Cryptology, P.O. Box 5159, Beijing, 100878, China}
\affiliation{School of Physics, University of Western Australia, Perth 6009, Australia}

\author{Fei Gao}
\email{gaof@bupt.edu.cn}
\affiliation{State Key Laboratory of Networking and Switching Technology, Beijing University of Posts and Telecommunications, Beijing 100876, China}

\author{Song Lin}
\affiliation{College of Mathematics and Informatics, Fujian Normal University, Fuzhou 350007, China}

\author{Jingbo Wang}
\email{jingbo.wang@uwa.edu.au}
\affiliation{School of Physics, University of Western Australia, Perth 6009, Australia}

\date{\today}

\begin{abstract}
Data compression can be achieved by reducing the dimensionality of high-dimensional but approximately low-rank datasets, which may in fact be described by the variation of a much smaller number of parameters. It often serves as a preprocessing step to surmount the curse of dimensionality and to gain efficiency, and thus it plays an important role in machine learning and data mining. In this paper, we present a quantum algorithm that compresses an exponentially large high-dimensional but approximately low-rank dataset in quantum parallel, by dimensionality reduction (DR) based on principal component analysis (PCA), the most popular classical DR algorithm. We show that the proposed algorithm achieves exponential speedup over the classical PCA algorithm when the original dataset are projected onto a polylogarithmically low-dimensional space. The compressed dataset can then be further processed to implement other tasks of interest, with significantly less quantum resources. As examples, we apply this algorithm to reduce data dimensionality for two important quantum machine learning algorithms, quantum support vector machine
and quantum linear regression for prediction.
This work demonstrates that quantum machine learning can be released from the curse of dimensionality to solve problems of practical importance.
\begin{description}
\item[PACS numbers]
03.67.Dd, 03.67.Hk
\end{description}
\end{abstract}

\maketitle

\section{Introduction}

In recent years, quantum machine learning (QML), an emerging interdisciplinary area that combines both quantum physics and machine learning, has become a booming research field attracting worldwide attention. Apart from applying classical machine learning methods to certain tasks of analysing quantum systems, such as quantum state separability classification \cite{MY18,Luetal18}, Hamiltonian learning \cite{WGFCA14,WGFCL14,Wangetal17} and characterization of an unknown unitary operation \cite{Betal10,Betal14}, the study of QML focuses more on designing quantum algorithms that can solve machine learning problems faster than the classical methods. As of now, and especially since the proposal of quantum linear system algorithm by Harrow et al. (HHL) \cite{HHL09}, a variety of quantum algorithms have been proposed to tackle various machine learning tasks, such as data classification \cite{LMR13,QSVM14,CD16,SFP17,DYLL17,SBSW18,SP18}, linear regression \cite{WBL12,SSP16,Wang17,QRR17,Yuetal18}, clustering analysis \cite{LMR13,ABG13}, association rules mining \cite{QARM16}, anomaly detection \cite{LR18}, and so on, achieving significant speedup over their classical counterparts. Moreover, some of them have been implemented experimentally \cite{Caietal15,Lietal15}. Overviews on QML can be seen in \cite{DB18,QML17}. Despite these achievements, QML has paid little attention to the important issue of \emph{curse of dimensionality} that might cause overfitting and degrade the performance of QML algorithms when processing high-dimensional datasets (sets of data points) \cite{PRML06,HML17}.

In the era of big data, a large high-dimensional dataset often lies in a lower-dimensional space. To reveal the intrinsic structure and mitigate the effects of the curse of dimensionality \cite{PRML06,HML17}, the technique of \emph{Dimensionality reduction} (DR), a process of reducing the number of features (dimensions) of a given dataset, but preserving the original dataset's information as well as possible, was put forward. Furthermore, since DR can be seen as the procedure of compressing
the original high-dimensional dataset into the low-dimensional dataset, which needs less resource in space and time to process,
DR in practice is often performed as a preprocessing step to support other machine learning tasks, such as classification and clustering analysis \cite{PRML06}. Generally, depending on whether the dimensionality is reduced by a linear or nonlinear method, DR is classified into two types: linear DR and nonlinear DR.  The most representative example of linear DR is \emph{principal component analysis} (PCA), which constructs a linear map that projects a high-dimensional dataset into a low-dimensional space, spanned by a few mutually orthogonal unit vectors, the \emph{principal components}, that maximally preserve the data variance \cite{PRML06,HML17}. Moreover, PCA mechanism is widely applied to a variety of both linear and nonlinear DR algorithms \cite{PRML06}, such as linear discriminant analysis (LDA) \cite{PRML06}, another important linear DR approach.

In the context of quantum computing, DR was first addressed in the PCA-based quantum state tomography algorithm by Lloyd et al.\cite{QPCA14}.
Given sufficient copies of a quantum state with density matrix $\rho$ encoding the correlation matrix of a dataset, the algorithm creates a
quantum state $\sigma=\sum_i r_i\ketbra{\chi_i}{\chi_i}\otimes \ketbra{\hat{r}_i}{\hat{r}_i}$, where $r_i$, $\ket{\chi_i}$, and $\hat{r}_i$ are respectively the eigenvalues,
corresponding eigenvectors (principal components), and the estimates of the eigenvalues of $\rho$. Then $\ket{\chi_i}$ and $\hat{r}_i$ can be
obtained by sampling the eigenvalue estimate register of $\sigma$ and this procedure will be highly efficient when $\rho$ is approximately low-rank \cite{QPCA14}.
For a more general Hermitian matrix $\mathcal{H}$ with eigenvalues $\lambda_i$ and corresponding eigenvectors $\ket{\varphi_i}$, Daskin \cite{Daskin16} subsequently proposed
another PCA-based quantum algorithm that uses amplitude amplification \cite{AA02} to obtain a linear combination of eigenvectors $\ket{\varphi_i}$, together with $\lambda_i$ ,
$\sum_{a\leq \lambda_i\leq b}\alpha_i\ket{\lambda_i}\ket{\varphi_i}$, where the eigenvalues $\lambda_i$ lie in a range $[a,b]$ and $\alpha_i$ are
the coefficients dependent on $\ket{\varphi_i}$. Later, Cong and Duan~\cite{CD16} suggested a quantum LDA algorithm for DR, which is similar to Llyod's algorithm \cite{QPCA14} and also yields a quantum state as $\sigma$,
but the state $\rho$ involved encodes some scatter matrices of a dataset.
However, all of these quantum algorithms only output the quantum basis states ($\ket{\chi_i}$ or $\ket{\varphi_i}$), that can be used to span
the low-dimensional space.  In other words, these quantum algorithms have not provided the desired quantum data compression, namely mapping the high-dimensional dataset into the low-dimensional space and
obtaining its corresponding low-dimensional dataset.  An intuitive idea to further complete this task based on these algorithms is to estimate the low-dimensional representation of
each high-dimensional quantum data point by suitably performing swap tests \cite{QSVM14,SSP16,BCWW01} between it and the quantum basis states.
Let us take Lloyd et al.'s algorithm \cite{QPCA14} as an example. For an arbitrary original data point prepared in a quantum state $\ket{\mathbf{x}}$, its corresponding low-dimensional
data can be represented, for some low dimension $d$, by the vector $\mathbf{y}=(\braket{\chi_1}{\mathbf{x}},\cdots,\braket{\chi_d}{\mathbf{x}})$, and each element
$\braket{\chi_i}{\mathbf{x}}$ can be estimated by swap tests \cite{QSVM14,SSP16,BCWW01}; note that $\braket{\chi_i}{\mathbf{x}}$ can be assumed to be non-negative
because both $\ket{\chi_i}$ and $-\ket{\chi_i}$ are eigenvectors (principal components).
However, it would be extremely exhausting to do so over an exponentially large high-dimensional dataset, and moreover, the low-dimensional classical data points cannot be directly applied to QML algorithms \cite{DB18,QML17}, which commonly require data points to be fed in quantum parallel. Therefore, it is important to design a quantum algorithm that can compress an exponentially large high-dimensional dataset by reducing its dimensionality and can be well adapted directly to QML to support other QML algorithms.

In this paper, we present a quantum algorithm for compressing big data by PCA, which is shown to be exponentially faster than the classical PCA when the data mostly lie in a polylogarithmically low-dimensional subspace of the original space.
Given a large set of high-dimensional data points, it maps them into a polylogarithmically low-dimensional space in quantum parallel by PCA.
This algorithm in fact can be seen as
a PCA-based quantum data compression scheme that compresses a register of qubits storing the original high-dimensional dataset
into significantly fewer qubits storing the compressed low-dimensional dataset. It is worth noting that this algorithm is different from the existing quantum data
compression schemes that compress an ensemble of identically prepared pure or mixed qubit states \cite{RMHetal14,YCH16,YCE16}, while our method compresses a sequence of different or even entangled qubit states.  The compressed data can then be used further to implement other tasks of interest and can be processed
with significantly less quantum resources in terms of the number of qubits and the number of quantum logic gates.
As examples, we also show that our algorithm can be applied to reduce the data dimensionality for two well-known quantum machine learning algorithms,
quantum support vector machine \cite{QSVM14} and quantum linear regression for prediction \cite{SSP16}.
This demonstrates that our algorithm has the potential to release QML from the curse of dimensionality.

The paper is organized as follows. Sec. \ref{Sec:QPCA} describes our quantum data compression algorithm by PCA, and analyses its runtime. In Sec. \ref{Sec:Discussions}, we discuss the adaptation of our algorithm to the existing QML algorithms. Conclusions are drawn in the last section.

\section{Quantum data compression by PCA}
\label{Sec:QPCA}

PCA is by far the most popular DR algorithm \cite{PRML06,HML17}. Given a dataset $\{\mathbf{x}_i\}_{i=1}^N$ that has $N$ data points and each one is represented by a $D$-dimensional column vector $\mathbf{x}_i=(x_{i1},x_{i2},\cdots,x_{iD})^T \in \mathbb{R}^D$, the dataset can be represented by a $N\times D$ matrix $X=(\mathbf{x}_1,\cdots,\mathbf{x}_N)^T$. PCA projects the dataset onto a low-dimensional space that can maximally preserve the data variance.

If we take the singular value decomposition form, i.e.,
\begin{eqnarray}
\label{eq:SVDX}
X=\sum_{j=1}^D \sigma_j\ket{\mathbf{u}_j}\bra{\mathbf{v}_j},
\end{eqnarray}
$\{\sigma_j\in \mathbb{R}_{\geq 0}\}_{j=1}^D$, $\{\ket{\mathbf{u}_j}\in \mathbb{R}^N\}_{j=1}^D$ and $\{\ket{\mathbf{v}_j}\in \mathbb{R}^D\}_{j=1}^D$ are, respectively, its singular values in descending order, the left singular vectors, and the right singular vectors being the principal components \cite{HML17}.

The original dataset can then be projected to a low-dimensional space of dimension $d$, which is spanned by the first $d$ principal components, $\ket{\mathbf{v}_1}, \ket{\mathbf{v}_2},\cdots,\ket{\mathbf{v}_d}$. The value of $d$ is commonly taken as the minimum $s$, such that the first $s$ principal components account for an accumulated variance larger than a pre-specified threshold $\vartheta$ that is close to 1, e.g., 0.95, that is
\begin{eqnarray}
\label{eq:dChoice}
d=\min_{s}\left\{s :\frac{\sum_{j=1}^s\sigma_j^2}{\sum_{j=1}^D\sigma_j^2} \geq \vartheta\right\}.
\end{eqnarray}
The projection is a linear map (operator) denoted by $V_d=(\ket{\mathbf{v}_1}, \ket{\mathbf{v}_2},\cdots,\ket{\mathbf{v}_d})$ and the new dataset is obtained by
\begin{eqnarray}
\label{eq:ProjVd}
Y&=&XV_d \\
\label{eq:SVDY}
&=& \sum_{j=1}^d \sigma_j\ket{\mathbf{u}_j}\bra{j}.
\end{eqnarray}
Here the $i$th row of the $N\times d$ matrix $Y$ is the transpose of the column vector
\begin{eqnarray}
\label{eq:Projyi}
\mathbf{y}_i=V_d^T\mathbf{x}_i,
\end{eqnarray}
where $\mathbf{y}_i=(y_{i1},y_{i2},\cdots,y_{id})^T \in \mathbb{R}^d$ and corresponds to the $i$th resultant $d$-dimensional data point after projecting $\mathbf{x}_i$ into the $d$-dimensional space.

In the context of quantum computing, all data points will be mapped in quantum parallel. The corresponding quantum algorithm transforms the quantum state (denoted by $\ket{\psi_s}$) storing the original data points into another quantum state (denoted by $\ket{\psi_e}$) storing the new low-dimensional data points, in quantum parallel:
\begin{eqnarray}
\label{map:qPCA}
\ket{\psi_s}:=\frac{\sum_{i=1}^N\ket{i}\otimes \mathbf{x}_i}{\norm{X}_F}\mapsto \ket{\psi_e}:=\frac{\sum_{i=1}^N\ket{i}\otimes \mathbf{y}_i}{\norm{Y}_F},
\end{eqnarray}
where $\norm{X}_F$ and $\norm{Y}_F$ are the Frobenius norms of $X$ and $Y$, and
\begin{eqnarray}
\label{eq:xi}
\mathbf{x}_i=\sum_{j=1}^D x_{ij}\ket{j}=\norm{\mathbf{x}_i}\ket{\mathbf{x}_i},\\
\label{eq:yi}
\mathbf{y}_i=\sum_{j=1}^d y_{ij}\ket{j}=\norm{\mathbf{y}_i}\ket{\mathbf{y}_i}.
\end{eqnarray}

Firstly, we assume the dataset $\{\mathbf{x}_i\}_{i=1}^N$, also represented by the matrix $X$, are stored in a quantum
random access memory (QRAM) \cite{QRAM08} with a suitable data structure, where sums of selected subsets of $\{\abs{X_{ij}}^2: i=1,2,\cdots,N,j=1,2,\cdots,D\}$ are subtly stored in binary trees \cite{KP16,WZP18}. The details about this structure can be found in its initial proposal for designing quantum recommendation systems \cite{KP16}, as well as its successful application in designing the quantum linear system algorithm for dense matrices \cite{WZP18}. Moreover, the structure allows us to efficiently perform the following two unitary operations in $O(\polylog(ND))$ time \cite{KP16,WZP18}.
\begin{eqnarray}
\label{eq:UMN}
&U_{\mathcal{M}}:& \ket{i}\ket{0}\mapsto\frac{\sum_{j=1}^D x_{ij}\ket{i}\ket{j}}{\norm{\mathbf{x}_i}},\\
&U_{\mathcal{N}}:& \ket{0}\ket{j}\mapsto\frac{\sum_{i=1}^N \norm{\mathbf{x}_i}\ket{i}\ket{j}}{\norm{X}_F}.
\end{eqnarray}
The operation $U_{\mathcal{M}}$ allows the preparation of a quantum state encoding an original data point, while $U_{\mathcal{N}}$ acts on the first register to encode the norm of the data points in the amplitudes. If we start with a quantum state $\ket{0}\ket{0}$ with a suitable number of qubits, we can use these two unitary operations to generate the desired initial state
\begin{eqnarray}
\label{state:psis}
\ket{\psi_s}
&=&U_{\mathcal{M}}U_{\mathcal{N}}\ket{0}\ket{0}\nonumber\\
&=&U_{\mathcal{M}}\frac{\sum_{i=1}^N \norm{\mathbf{x}_i}\ket{i}\ket{0}}{\norm{X}_F}\nonumber\\
&=&\frac{\sum_{i=1}^N \sum_{j=1}^D x_{ij}\ket{i}\ket{j}}{\norm{X}_F}.
\end{eqnarray}
Therefore, thanks to this data structure in QRAM, $\ket{\psi_s}$ can be efficiently generated in $O(\polylog(ND))$ time.

\subsection{Algorithm}

To generate the final desired state $\ket{\psi_e}$ in the transformation of Eq.~(\ref{map:qPCA}), our algorithm works as the following steps.

(1) \textbf{Extract quantum principal components}. This step aims to obtain the first $d$ principal components in the quantum state form, $\ket{\mathbf{v}_1}, \ket{\mathbf{v}_2},\cdots,\ket{\mathbf{v}_d}$. The corresponding variance proportions they account for are
\begin{eqnarray}
\label{eq:VarPro}
\lambda_j:=\frac{\sigma_j^2}{\sum_{j=1}^D \sigma_j^2}, j=1,2,\cdots,D.
\end{eqnarray}
According to the singular value decomposition form (Eq.~\eqref{eq:SVDX}) of $X$, $\ket{\psi_s}$ (Eq.~\eqref{state:psis}) can be rewritten as
\begin{eqnarray}
\label{state:SVDX}
\ket{\psi_s}=\sum_{j=1}^D \sqrt{\lambda_j} \ket{\mathbf{u}_j}\ket{\mathbf{v}_j}.
\end{eqnarray}
The second register is in the state
\begin{eqnarray}
\rho&=&\Tr_1(\ket{\psi_s})\nonumber\\
&=&\sum_{j=1}^D \lambda_j \ket{\mathbf{v}_j}\bra{\mathbf{v}_j},
\end{eqnarray}
which is actually equal to $X^TX/\Tr(X^TX)$.

Based on the density matrix exponentiation technique, which takes a number of copies of $\rho$ to apply $e^{-i\rho t}$ (for some time $t$) \cite{QPCA14}, we perform phase estimation on the second register of $\ket{\psi_s}$, and obtain the state
\begin{eqnarray}
\label{state:XPE}
\sum_{j=1}^D \sqrt{\lambda_j} \ket{\mathbf{u}_j}\ket{\mathbf{v}_j}\ket{\lambda_j}.
\end{eqnarray}
Then we sample this state by measuring the last register to reveal the eigenvalues $\lambda_j$, and obtain the corresponding states $\ket{\mathbf{u}_j}$ and $\ket{\mathbf{v}_j}$. Apparently, $\lambda_j$ can be revealed with probability $\lambda_j$. Since the sum $\sum_{j=1}^d\lambda_j\ge \vartheta \approx 1$, sampling the state of Eq.~\eqref{state:XPE} for $O(d)$ times suffices to obtain $\lambda_j$ if each $\lambda_j$ scales as $O(1/d)$ .

(2) \textbf{Introduce the anchor state}. We randomly pick out one original data point $\mathbf{x}\in \{\mathbf{x}_i\}_{i=1}^N$ and, via an unitary operation $U_\mathbf{x}:\ket{0}\mapsto \ket{\mathbf{x}}$, prepare its corresponding quantum state $\ket{\mathbf{x}}$, which we call the \emph{anchor state}. It is easy to see that $U_\mathbf{x}$ can be readily implemented via $U_{\mathcal{M}}$ (Eq.~\eqref{eq:UMN}) and thus can be efficiently implemented in $O(\polylog D)$ time.

Since $\{\ket{\mathbf{v}_j}\}_{j=1}^D$ constitutes a basis for the space $\mathbb{R}^D$, $\ket{\mathbf{x}}$ can be written as a linear combination of these basis states, that is,
\begin{eqnarray}
\label{state:x}
\ket{\mathbf{x}}=\beta_1\ket{\mathbf{v}_1}+\beta_2\ket{\mathbf{v}_2}+\cdots+\beta_D\ket{\mathbf{v}_D},
\end{eqnarray}
where $\beta_j=\bra{\mathbf{v}_j}\ket{\mathbf{x}}$. Without loss of the generality,  we assume every $\beta_j\ge 0$; if $\beta_j<0$, we can replace $\ket{\mathbf{v}_j}$ with $-\ket{\mathbf{v}_j}$ so that $\beta_j=\abs{\beta_j}\ge 0$. Note that, just as for $\ket{\mathbf{v}_j}$, $-\ket{\mathbf{v}_j}$ is also the eigenvector of $X^TX$ with eigenvalue $\lambda_j$ and it can also be seen as a principal component.

As the original dataset $\{\mathbf{x}_i\}_{i=1}^N$ approximately lie in the subspace spanned by the first $d$ principal components $\{\ket{v_j}\}_{j=1}^d$, $\ket{\mathbf{x}}$ will with high probability have a large support in this subspace. This means that $\sum_{j=1}^d \beta_j^2\approx 1$ and $\beta_j^2=O(1/d)$. The value of $\beta_j^2$ can be estimated to accuracy $O(\epsilon_{\beta}/\sqrt{d})$, or equivalently, $\beta_j$ can be estimated to accuracy $O(\epsilon_{\beta})$, with $O\left(\beta_j^2(1-\beta_j^2)d/\epsilon_{\beta}^2\right)=O(1/\epsilon_{\beta}^2)$ swap tests \cite{QSVM14,SSP16,BCWW01}.

(3) \textbf{Projection}. This step is to generate the final desired state $\ket{\psi_e}$. To understand the basic idea of transforming $\ket{\psi_s}$ to $\ket{\psi_e}$, we rewrite any original data point $\mathbf{x}_i$ in the basis $\{\ket{\mathbf{v}_1},\ket{\mathbf{v}_2},\cdots,\ket{\mathbf{v}_D}\}$ as
\begin{eqnarray*}
\mathbf{x}_i=(\sum_{j=1}^D \ket{\mathbf{v}_j}\bra{\mathbf{v}_j})\mathbf{x}_i=\sum_{j=1}^D y_{ij}\ket{\mathbf{v}_j}.
\end{eqnarray*}
This means that $\ket{\psi_s}$ can be mathematically reformulated as
\begin{eqnarray}
\label{state:NewX}
\ket{\psi_s}=\frac{\sum_{i=1}^N\ket{i}\otimes \sum_{j=1}^D y_{ij}\ket{\mathbf{v}_j}}{\norm{X}_F},
\end{eqnarray}
and further implies that $\ket{\psi_e}$ can be obtained if we can perform the mapping: $\ket{\mathbf{v}_j}\mapsto \ket{j}$ on $\ket{\psi_s}$ and then truncate it to keep the first $d$ terms. Based on this idea, the projection is achieved as follows.

(3.1) \textbf{Phase estimation}. Again use the ability to apply $e^{-i\rho t}$ to perform phase estimation on the second register of the state of Eq.~\eqref{state:NewX}, and obtain the state
\begin{eqnarray}
\label{state:NewXPE}
\frac{\sum_{i=1}^N\sum_{j=1}^D y_{ij}\ket{i}\ket{\mathbf{v}_j}\ket{\lambda_j}}{\norm{X}_F},
\end{eqnarray}
which is mathematically equal to the state of Eq.~\eqref{state:XPE}.

\begin{figure}[htbp]
\includegraphics[scale=0.6]{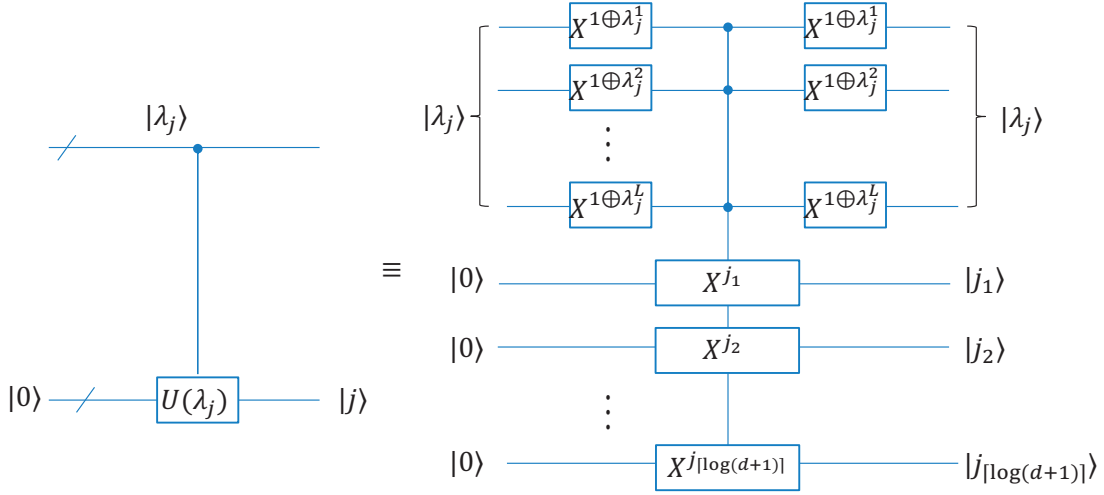}
\caption{\label{fig:CUJ} Quantum circuit for $CU_j$, where the upper register stores the eigenvalues $\ket{\lambda_j}$ and the lower one is the index register. Here / denotes a bundle of qubits, and $X$ denotes quantum NOT gate. $X^{1\oplus \lambda_{j}^k}=I(X)$ if $\lambda_{j}^k=1(0)$ for $k=1,2,\cdots,L$, and $X^{j_l}=I(X)$ if $j_l=0(1)$ for $l=1,2,\cdots,\lceil \log(d+1)\rceil$. The circuit implements the map $\ket{0}\mapsto \ket{j} \equiv \ket{j_1j_2\cdots j_{\lceil \log(d+1) \rceil}}$ on the index register if the eigenvalue register is in the state $\ket{\lambda_j} \equiv \ket{\lambda_j^1\lambda_j^2\cdots\lambda_j^L}$, and otherwise does nothing.}
\end{figure}

(3.2) \textbf{Append an index register}. We append an \emph{index register} of $\lceil \log(d+1) \rceil$ qubits in the state $\ket{0}$, perform $d$ controlled unitary operations, $CU(\lambda_1),CU(\lambda_2),\cdots,CU(\lambda_d)$, where
\begin{eqnarray}
\label{eq:CUJ}
CU(\lambda_j): \ket{\lambda_j}\ket{0}\mapsto \ket{\lambda_j}\ket{j},
\end{eqnarray}
and then obtain the state
\begin{eqnarray}
\label{state:IndexReg}
\frac{\sum_{i=1}^N y_{ij}\ket{i}\ket{\mathbf{v}_j}\ket{\lambda_j}(\sum_{j=1}^d \ket{j}+\sum_{j=d+1}^D\ket{0})}{\norm{X}_F}.
\end{eqnarray}
Each controlled unitary operation $CU_j$ can be implemented efficiently because the eigenvalues $\lambda_j$ have been revealed in step (1). Assuming $L$ qubits are used to store the eigenvalues $\lambda_j$ ($j=1,2,\cdots,d$), and $\lambda_j$ and $j$ respectively have the binary representation $\lambda_j=\lambda_j^1\lambda_j^2\cdots \lambda_j^L$ and $j=j_1j_2\cdots j_{\lceil \log(d+1) \rceil}$, the quantum circuit for $CU_j$ is shown in Fig.~\ref{fig:CUJ}.

(3.3) \textbf{Inverse phase estimation}. We undo step (3.1), discard the eigenvalue register, and have the state
\begin{eqnarray}
\label{state:IPE}
\frac{\sum_{i=1}^N y_{ij}\ket{i}\ket{\mathbf{v}_j}(\sum_{j=1}^d \ket{j}+\sum_{j=d+1}^D\ket{0})}{\norm{X}_F}.
\end{eqnarray}

(3.4) \textbf{Controlled rotation}. Append another qubit in the state $\ket{0}$ as the last register, and perform $d$ controlled unitary rotation operations $CR(\beta_1),CR(\beta_2),\cdots,CR(\beta_d)$, conditioned on $\ket{j}$, where
\begin{eqnarray}
\label{eq:CRJ}
CR(\beta_j): \ket{j}\ket{0}\mapsto \ket{j}\left(\frac{C}{\hat{\beta}_j}\ket{1}+\sqrt{1-\frac{C^2}{\hat{\beta}_j^2}}\ket{0}\right),
\end{eqnarray}
for $j=1,2,\cdots,d$. Here $\hat{\beta}_j$ denotes the estimate of $\beta_j$ obtained in step (2) and $C=O(\min_j \hat{\beta_j})=O(1/\sqrt{d})$. After rotation, we have the state of the whole system,
\begin{eqnarray}
\label{state:XCR}
\sum_{i=1}^N \frac{y_{ij}}{\norm{X}_F}\ket{i}\ket{\mathbf{v}_j}\left(\sum_{j=1}^d \ket{j}\Big(\frac{C}{\hat{\beta}_j}\ket{1}+\sqrt{1-\frac{C^2}{\hat{\beta}_j^2}}\ket{0}\Big)+\sum_{j=d+1}^D\ket{0}\ket{0}\right).
\end{eqnarray}

It should be noted that, since the classical information of each $\hat{\beta}_j$ is provided, each controlled rotation $CR(\beta_j)$ can be directly constructed and efficiently implemented. This is much more resource-saving than the controlled rotations which are widely used in the HHL-based quantum algorithms \cite{HHL09,CD16,DYLL17,WBL12,SSP16,QRR17,Yuetal18}, but need quantum arithmetic circuits \cite{HRS18} taking substantially more qubits and runtime.

(3.5) \textbf{Projective measurement}. Measure the second register of $\ket{\mathbf{v}_j}$ to see whether it is in the state $\ket{\mathbf{x}}=U_{\mathbf{x}}\ket{0}$ by performing $U_{\mathbf{x}}^{-1}$ followed by the projective measurement $\ket{0}\bra{0}$. At the same time, measure the last qubit to see whether it is in the state $\ket{1}$ by performing the projective measurement $\ket{1}\bra{1}$.  If both measurements succeed, we trace out these two registers and have the state of the other two registers
\begin{eqnarray}
\label{state:psie}
\frac{\sum\limits_{i=1}^N\sum\limits_{j=1}^d \frac{y_{ij}}{\norm{X}_F}\frac{C\beta_j}{\hat{\beta}_j}\ket{i}\ket{j}}{\sqrt{\sum\limits_{i=1}^N\sum\limits_{j=1}^d\left(\frac{y_{ij}}{\norm{X}_F}\frac{C\beta_j}{\hat{\beta}_j}\right)^2}}
&\approx& \frac{\sum\limits_{i=1}^N\sum\limits_{j=1}^d y_{ij}\ket{i}\ket{j}}{\sqrt{\sum\limits_{i=1}^N\sum\limits_{j=1}^d y_{ij}^2}}.
\end{eqnarray}
The approximation holds because $\beta_j\approx \hat{\beta}_j$ as shown in step (2), and the right-hand side is exactly $\ket{\psi_e}$ by plugging Eq.~\eqref{eq:yi} into the right-hand side of Eq.~\eqref{map:qPCA}. This means that we finally obtain a state $\approx \ket{\psi_e}$ storing the new low-dimensional dataset $\{\mathbf{y}_i\}_{i=1}^N$ in quantum parallel, and thus complete the transformation of Eq.~\eqref{map:qPCA} we desire. Due to the linearity of the entire process, our algorithm can reduce the dimensionality of only a subset of the original data points,
\begin{eqnarray}
\label{map:subsetqPCA}
\frac{\sum_{i\in S}\ket{i}\otimes \mathbf{x}_i}{\sqrt{\sum_{i\in S}\norm{\mathbf{x}_i}^2}}\mapsto  \frac{\sum_{i\in S}\ket{i}\otimes \mathbf{y}_i}{\sum_{i\in S}\sqrt{\norm{\mathbf{y}_i}^2}},
\end{eqnarray}
where $S \subseteq \{1,2,\cdots,N\}$.
Furthermore, for each individual data point prepared in the quantum state $\ket{\mathbf{x}_i}$, it is also straight forward to see that our algorithm implements
\begin{eqnarray}
\label{map:singleqPCA}
\ket{\mathbf{x}_i} \mapsto \ket{\mathbf{y}_i},
\end{eqnarray}
without introducing the first register $\ket{i}$.

\begin{figure}[htb]
\includegraphics[scale=0.5]{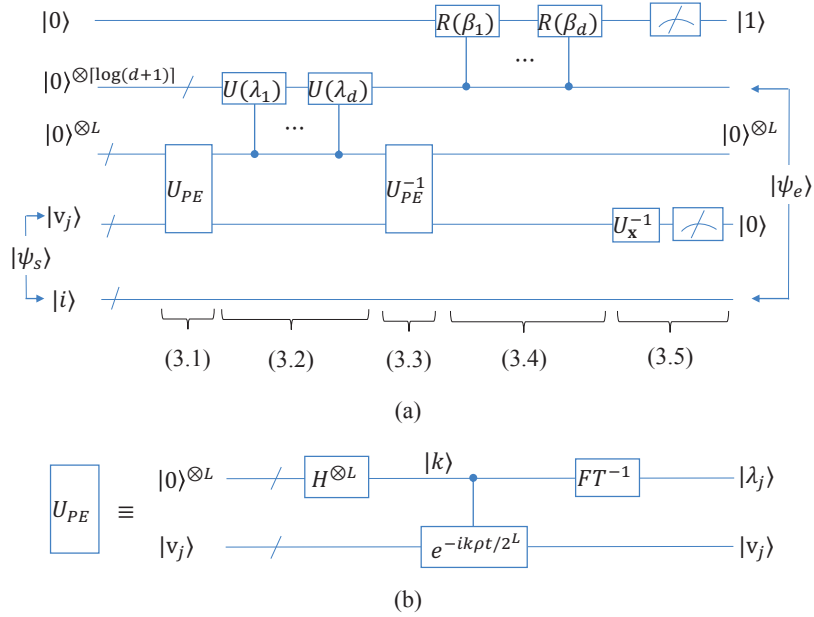}
\caption{\label{fig:QCirprojection} (a) Quantum circuit for step (3), where the unitary operation $U_{PE}$ represents the phase estimation algorithm and its quantum circuit is depicted in (b). The unitary operations controlled-$U(\lambda_1),\cdots,U(\lambda_d)$ and controlled-$R(\beta_1),\cdots,R(\lambda_d)$ correspond with $CU(\lambda_1),\cdots,CU(\lambda_d)$ and $CR(\beta_1),\cdots,CR(\beta_d)$, respectively. The five substeps of step (3) are marked as $(3.1),(3.2),\cdots,(3.5)$. (b) The quantum circuit for $U_{PE}$, where the unitary operations $H$ and $FT$ denotes the Hadamard gate and the quantum Fourier transformation \cite{QCQI10}, resepctively.}
\end{figure}

Now, we have finished describing the whole algorithm. The first two steps can be seen as supports to step (3), the key step in our algorithm. The quantum circuit for step (3) is shown in Fig. \ref{fig:QCirprojection}

\subsection{Runtime analysis}

In step (1), according to \cite{QPCA14}, taking $O(1/\epsilon_\lambda^3)$ copies of $\rho$ to implement the phase estimation algorithm \cite{QCQI10} ensures that the eigenvalues $\lambda_j$ can be estimated within error $\epsilon_\lambda$. Since preparing each copy of $\rho$ needs preparation of one copy of $\ket{\psi_s}$ with time complexity $O(\polylog(ND))$, generating the state of Eq.~\eqref{state:XPE} takes runtime $O\left(\polylog(ND)/\epsilon_\lambda^3\right)$. In addition, as mentioned in step (1), sampling the state of Eq.~\eqref{state:XPE} for $O(d)$ times guarantees that, with high probability, we can obtain each $\lambda_j$ and one copy of each $\ket{\mathbf{v}_j}$.

In step (2), based on the natural assumption that $\beta_j=O(1/\sqrt{d})$ for $j=1,2, \cdots, d$, each $\beta_j$ can be estimated within error $O(\epsilon_\beta)$ with $O(1/\epsilon_\beta^2)$ swap tests \cite{QSVM14,SSP16,BCWW01}, which takes $O(1/\epsilon_\beta^2)$ copies of $\ket{\mathbf{x}}$ as well as $\ket{\mathbf{v}_j}$. Since generating one copy of $\ket{\mathbf{x}}$ takes $O(\polylog D)$ time and each swap test takes $O(\log D)$ elementary gates, the total runtime for step (2) is $O(d\polylog D/\epsilon_\beta^2)$ under the condition that those copies of $\ket{\mathbf{v}_j}$ are provided by step (1). For step (1), the requirement of $O(1/\epsilon_\beta^2)$ copies of $\ket{\mathbf{v}_j}$ implies the overall runtime
\begin{eqnarray*}
O\left(\frac{d\polylog(ND)}{\epsilon_\beta^2\epsilon_\lambda^3}\right).
\end{eqnarray*}

The runtime for steps (3.1) and (3.3) obviously coincides with that for the procedure of generating the state of Eq.~\eqref{state:XPE} in step (1). In step (3.2), each $CU_j$ for $j=1,2,\cdots,d$ takes $O(\log(1/\epsilon_\lambda)\log(d+1))$ elementary gates \cite{QCQI10,QGate95}, so step (3.2) takes $O\left(d\log(1/\epsilon_\lambda)\log(d+1)\right)$ time. According to \cite{QGate95}, step (3.4) generally takes $O(d\log(d+1))$ time. In step (3.5), just as $U_{\mathbf{x}}$ in step (2), the unitary operation $U_{\mathbf{x}}^{-1}$ also runs in time $O\left(\polylog D\right)$. Finally, the success probability of measurement in step (3.5) is
\begin{eqnarray}
\label{eq:succprob}
p: &=& \sum_{i=1}^N\sum_{j=1}^d\left(\frac{y_{ij}}{\norm{X}_F}\frac{C\beta_j}{\hat{\beta}_j}\right)^2 \nonumber\\
&\approx& \frac{C^2\norm{Y}_F^2}{\norm{X}_F^2}\nonumber\\
&=&O(1/d),
\end{eqnarray}
because $C=O(1/\sqrt{d})$, and
\begin{eqnarray*}
\frac{\norm{Y}_F^2}{\norm{X}_F^2}=\frac{\sum_{j=1}^d\sigma_j^2}{\sum_{j=1}^D\sigma_j^2} \geq  \vartheta=O(1),
\end{eqnarray*}
according to Eqs.~\eqref{eq:SVDX} and \eqref{eq:SVDY}. That is to say, $O(d)$ measurements are required to obtain $\ket{\psi_e}$ with high probability; the number of repetitions can be reduced to $O(\sqrt{d})$ via amplitude amplification \cite{AA02}.

\begin{table}[htb]
\caption{\label{tab:steptimecomplexity}%
The time complexity of each step of our algorithm. Here the $\sqrt{d}$ factor occurring in each substep of step (3) corresponds to the number of repetitions for amplitude amplification in step (3.5).}
\begin{ruledtabular}
\begin{tabular}{cc}
Steps &Time complexity \\ \hline
(1) &$O(\epsilon_\beta^{-2}\epsilon_\lambda^{-3} d\polylog(ND))$ \\
(2) &$O(d\polylog D/\epsilon_\beta^2)$ \\
(3.1) & $O(\sqrt{d}\cdot\polylog(ND)/\epsilon_\lambda^3)$ \\
(3.2) & $O(\sqrt{d}\cdot d\log(1/\epsilon_\lambda)\log(d+1))$ \\
(3.3) & $O(\sqrt{d}\cdot\polylog(ND)/\epsilon_\lambda^3)$ \\
(3.4) & $O(\sqrt{d}\cdot d\log(d+1))$\\
(3.5) & $O(\sqrt{d}\cdot\polylog D)$
\end{tabular}
\end{ruledtabular}
\end{table}

Moreover, according to the analysis of the HHL algorithm \cite{HHL09}, the final error for generating $\ket{\psi_e}$ mainly comes from the inverse of $\beta_j$. Since $\beta_j$ is estimated with error $O(\epsilon_\beta)$, the relative error of estimating $1/\beta_j$ is $O(\epsilon_\beta/\beta_j)=O(\epsilon_\beta/\sqrt{d})$ and thus the final state $\ket{\psi_e}$ induces error $O(\epsilon_\beta/\sqrt{d})$. To ensure the final error be within $\epsilon$, we should take $\epsilon_\beta=O(\sqrt{d}\epsilon)$.

The time complexity of each step of our algorithm is shown in the Table \ref{tab:steptimecomplexity}. Putting all runtime together and taking $\epsilon_\beta=O(\sqrt{d}\epsilon)$ as shown above, our algorithm has the overall runtime
\begin{eqnarray*}
O\left(\epsilon^{-2}\epsilon_\lambda^{-3}\polylog(ND)+d^{3/2}\log( 1/\epsilon_\lambda )\log(d+1)\right).
\end{eqnarray*}
In addition, if every $\lambda_j=O(1/d)$, $\epsilon_\lambda$ should take $O(1/d)$ and thus the overall runtime will be $O(d^3\polylog(ND)/\epsilon^2)$. This means that it has runtime $O(\polylog(N,D))$ when $d =O(\polylog(D))$, achieving exponential speedup over the classical PCA algorithm whose runtime is $O(\poly(N,D))$ \cite{PRML06,HML17}.

According to the above analysis and the transformation of Eq.~\eqref{map:singleqPCA}, it is easy to see that our algorithm can efficiently transform each quantum data point $\ket{\mathbf{x}_i}$ of the high-dimensional quantum dataset $\{\ket{\mathbf{x}_i}\}_{i=1}^N$ into a polylogarithmically lower-dimensional quantum data $\ket{\mathbf{y}_i}$ of polylogarithmically fewer qubits. This implies that, under certain conditions, our algorithm breaks the limits on quantum dimensionality reduction \cite{QDR15}. The result of \cite{QDR15} shows that the dimension of a set of high-dimensional quantum states cannot be significantly reduced while preserving the 2-norm distance of each pairwise states. This leaves an open question of whether significant dimension reduction is possible when multiple copies of the input states (whose dimensionality is to be reduced) are available \cite{QDR15}. Our algorithm gives an positive answer to this question, in the sense that, by taking take polylogarithmically fewer copies of the input states $\{\ket{\mathbf{x}_i}\}_{i=1}^N$, as shown above, our algorithm can map each state $\ket{\mathbf{x}_i}$ into the polylogarithmically lower-dimensional state $\ket{\mathbf{y}_i}$, while well preserving the 2-norm distances of most pairwise states. The 2-norm distances of most pairwise states are well preserved because our quantum algorithm is based on PCA which guarantees that $\braket{\mathbf{y}_{i_1}}{\mathbf{y}_{i_2}}\approx \braket{\mathbf{x}_{i_1}}{\mathbf{x}_{i_2}}$ for most $i_1,i_2\in \{1,2,\cdots,n\}$. Note that $\norm{\ket{a}-\ket{b}}_2=\sqrt{2(1-\braket{a}{b})}$ for any two same-dimensional quantum states with real amplitudes.

\section{Adaptation to quantum machine learning}
\label{Sec:Discussions}

In this section, we show that, to reduce the data dimensionality, our quantum algorithm can be applied to two well-known supervised quantum machine learning algorithms, quantum support vector machine (QSVM) \cite{QSVM14} and quantum linear regression for prediction \cite{SSP16}.

\subsection{Quantum support vector machine}
Support vector machine (SVM) is an important supervised machine learning algorithm for data classification. Given $N$ training data points $\{(\mathbf{x}_i,z_i):\mathbf{x}_i\in \mathbb{R}^D,z_i=\pm 1\}_{i=1}^N$, where $z_i$ identifies which class $\mathbf{x}_i$ belongs to, the task of SVM is to classify a vector into one of two classes.
The least-squares approximation of SVM is found by solving the following system of linear equations:
\begin{eqnarray}
\label{eq:LSSVM}
F\left(\begin{matrix}
a \\
\mathbf{b}
\end{matrix}\right)
=\left(
\begin{matrix}
0 & \mathbf{1} \\
\mathbf{1} &K+\gamma I
\end{matrix}
\right)
\left(
\begin{matrix}
a \\
\mathbf{b}
\end{matrix}
\right)
=
\left(
\begin{matrix}
0 \\
\mathbf{z}
\end{matrix}
\right).
\end{eqnarray}
Here, $\mathbf{1}=(1,\cdots,1)^T$, $\gamma \in \mathbb{R}$ is a constant, $K_{ij}=\mathbf{x}_i^T \mathbf{x}_j$; $a \in \mathbb{R}$ and $\mathbf{b}=(b_1,\cdots,b_N)$; and $\mathbf{z}=(z_1,\cdots,z_N)$. When obtaining $a$ and $\mathbf{b}$, one can predict the class of a new data point $\mathbf{x}_0$ by $z_0=\sgn(\sum_{j=1}^N b_j\mathbf{x}_0^T \mathbf{x}_j+a)$.

\begin{figure}[htb]
\includegraphics[scale=0.5]{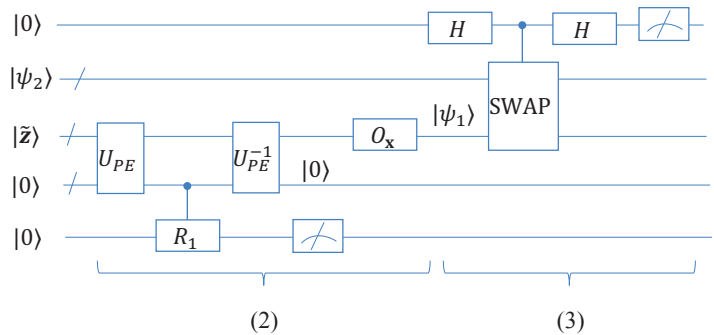}
\caption{\label{fig:QSVM} Schematic quantum circuit for QSVM, where $U_{PE}$ represents the phase estimation algorithm whose circuit is same as
Fig.~\ref{fig:QCirprojection} (b) except that the unitary operation controlled is now $e^{-i F/\Tr(F)t}$ for some $t$.
Here the state $\ket{\tilde{\mathbf{z}}}$ corresponds to the normalized right-hand side of Eq.~\eqref{eq:LSSVM}, controlled-$R_1$ denotes the controlled rotation required in step (2) for matrix inversion \cite{QSVM14}, and SWAP corresponds to the swap operation. The numbers (2) and (3) respectively mark steps (2) and (3).}
\end{figure}

Following the above approach, the QSVM algorithm \cite{QSVM14} can be briefly summarized by
the following steps and the corresponding schematic quantum circuit is shown in Fig.~\ref{fig:QSVM}.

(1) Assuming quantum oracles accessing the training data points, denoted by $O_\textbf{x}$, are provided, QSVM generates a number of copies of the quantum state $\ket{\chi_1}=\sum_{i=1}^N \ket{i}\otimes \mathbf{x}_i/\sqrt{\sum_{i=1}^N \norm{\mathbf{x}_i}^2}$.

(2) By using the quantum matrix insertion approach \cite{HHL09} together with exponentiation of the density matrix $K/\Tr(K)=\Tr_2(\ket{\chi_1}\bra{\chi_1})$ \cite{QPCA14}, QSVM solves the above linear system of equations (i.e., Eq.~\eqref{eq:LSSVM}), creates the quantum state $\ket{a,\mathbf{b}}=a\ket{0}+\sum_{j=1}^Nb_j\ket{j}$, and, by calling $O_\textbf{x}$, further generates
\begin{eqnarray*}
\label{state:psi1}
\ket{\psi_1}=\frac{a\ket{0}\otimes\ket{0}+\sum_{j=1}^Nb_j\ket{j}\otimes \mathbf{x}_j}{\sqrt{a^2+\sum_{j=1}^Nb_j^2\norm{\mathbf{x}_j}^2}}.
\end{eqnarray*}

(3) Prepare the state
$\ket{\psi_2}=(\ket{0}\otimes\ket{0}+\sum_{j=1}^N \ket{j}\otimes \mathbf{x_0})/\sqrt{1+N\norm{\mathbf{x_0}}^2}$,
and obtain $\bra{\psi_1}\psi_2\rangle$ via swap test \cite{QSVM14}, which reveals $z_0$, the predictive class of the data point $\mathbf{x}_0$.

Our quantum algorithm can be applied to QSVM to reduce the data dimensionality. Specifically, in step (1) our algorithm transforms $\ket{\chi_1}$ into another state $\ket{\chi_1'}=\sum_{i=1}^N \ket{i}\otimes \mathbf{y}_i/\norm{Y}_F$, where $\mathbf{y}_i$ is the low-dimensional data representation of $\mathbf{x}_i$ by PCA. In addition, step (2) will instead proceed with exponentiation of the density matrix $\Tr_2(\ket{\chi_1'}\bra{\chi_1'})$, which encodes the low-dimensional data points $\{\mathbf{y}_1,\cdots,\mathbf{y}_N\}$, and $\mathbf{x}_j$ in the state $\ket{\psi_1}$ will be replaced with $\mathbf{y}_j$. Finally, in step 3, $\mathbf{x}_0$ is projected to the low-dimensional space with the data point $\mathbf{y}_0$ in the state $\ket{\psi_2}$, and $\mathbf{y}_0$ can be successfully classified via swap tests.

\subsection{Quantum linear regression algorithm for prediction}

Linear regression is another important supervised learning task. Given $N$ training data points, for simplicity also denoted by $\{(\mathbf{x}_i,z_i):\mathbf{x}_i\in \mathbb{R}^D,z_i \in \mathbb{R}\}_{i=1}^N$, the objective is to construct a linear model $f(x)=\mathbf{x}^T\mathbf{w}$ that can best describe the linear relationship between $\mathbf{x}_i$ and $z_i$ and use this model to predict $z_0=f(\mathbf{x}_0)$ for a new input $\mathbf{x}_0$. Using the least-squares method, the optimal $\mathbf{w}$ is given by $\mathbf{w}=(X^TX)^{-1}X^T\mathbf{z}$, where $X=(\mathbf{x}_1,\cdots,\mathbf{x}_N)^T$ and $\mathbf{z}=(z_1,\cdots,z_N)$, and thus $z_0=\sum_{j=1}^D\mathbf{w}^T\mathbf{x}_0$. Taking the singular value decomposition form, $X=\sum_{j=1}^R \sigma_j\ket{\mathbf{u}_j}\bra{\mathbf{v}_j}$, where $R$ is the rank of $X$, $z_0=\sigma_j^{-1}\mathbf{x}_0^T \ket{\mathbf{v}_j}\bra{\mathbf{u}_j}\mathbf{z}$.

\begin{figure}[htb]
\includegraphics[scale=0.5]{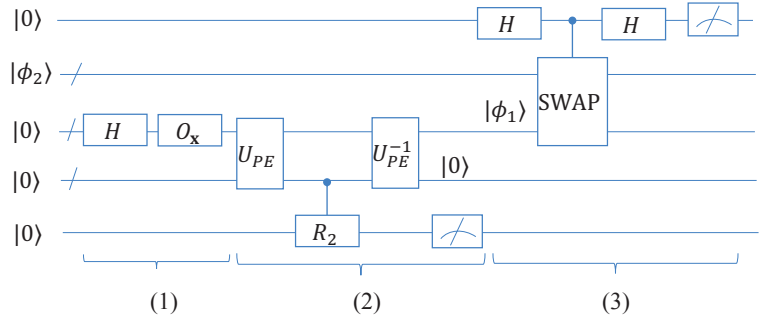}
\caption{\label{fig:QLR} Schematic quantum circuit for QLR, where $U_{PE}$ also represents the phase estimation algorithm (as
shown in Fig.~\ref{fig:QCirprojection} (b)) for the unitary operation $e^{-i\Tr_2(\ket{\chi_2}\bra{\chi_2})t}$ for some $t$.
 Also, the controlled-$R_2$ also denotes the controlled rotation involved in matrix inversion \cite{SSP16,HHL09} in step (2), and the numbers (1)-(3) correspond to the three steps.}
\end{figure}

In the context of quantum computing, Schuld et al. in 2016 proposed a quantum algorithm (QLR) \cite{SSP16} for implementing the above task, where, for simplicity, $\norm{X}_F$, $\norm{\mathbf{z}}$ and $\norm{\mathbf{x}_0}$ are all assumed to be 1. The algorithm can be outlined as follows and the corresponding schematic quantum circuit is depicted in Fig.~\ref{fig:QLR}.

(1) Assuming quantum oracles accessing the training data points, also denoted by $O_\textbf{x}$, are provided, QLR generates the initial quantum state $\ket{\chi_2}=\sum_{i=1}^N \mathbf{x}_i \otimes \ket{i}/\sqrt{N} =\sum_{j=1}^R\sigma_j\ket{\mathbf{v}_j}\ket{\mathbf{u}_j}$.

(2)Using the quantum matrix insertion approach \cite{HHL09} together with the exponentiation of the density matrix $\Tr_2(\ket{\chi_2}\bra{\chi_2})$ \cite{SSP16}, QLR transforms $\ket{\chi_2}$ into another state $\ket{\phi_1} \propto \sum_{j=1}^R 1/\sigma_j\ket{\mathbf{v}_j}\ket{\mathbf{u}_j}$.

(3) Prepare the state
$\ket{\phi_2}=\ket{\mathbf{x}_0}\ket{\mathbf{z}}$ with $\ket{\mathbf{z}}$ being the quantum-state form of (normalized) $\mathbf{z}$,
and then estimate $\bra{\phi_1}\phi_2\rangle \propto z_0$ up to some measurable factor \cite{SSP16}, via swap test \cite{QSVM14} with an ancilla qubit.

Our quantum algorithm can also be applied to QLR. In step (1), by our quantum algorithm, the state $\ket{\chi_2}$ is transformed into $\ket{\chi_2'} \propto \sum_{i=1}^N \mathbf{y}_i\otimes \ket{i}$, where $\mathbf{y}_i$ is the low-dimensional data corresponding to $\mathbf{x}_i$. Step (2) will work with the density matrix $Tr_2(\ket{\chi_2'}\bra{\chi_2'})$, which encodes the low-dimensional data points $\{\mathbf{y}_1,\cdots,\mathbf{y}_N\}$, and transform $\ket{\chi_2'}$ into $\ket{\phi_1'}$. In the last step, we by our algorithm generate the state $\ket{\phi_2'}:=\ket{\mathbf{y}_0}\ket{\mathbf{z}}$ from $\ket{\phi_2}$, where $\ket{\mathbf{y}_0}$ is the normalized quantum state representation of low-dimensional $\mathbf{x}_0$. Just as predicting $\mathbf{x}_0$, $\ket{\phi_2'}$ is then taken into swap tests with $\ket{\phi_1'}$ to predict the target of $\ket{\mathbf{y}_0}$.

In addition to the above two examples, it is promising that, to reduce the data dimensionality and avoid the curse of dimensionality, our algorithm could also be applied to other QML algorithms, where the states as $\ket{\psi_s}$, storing the data points in quantum parallel, are involved.

\section{Conclusions}
\label{Sec:Conculsions}
We have presented a quantum algorithm that compresses a large high-dimensional dataset in quantum parallel by reducing their dimensionality based on PCA. The algorithm is shown to be exponentially faster than the classical PCA algorithm, when the compressed dataset is of polylogarithmically low dimensionality. The compressed dataset can then be further processed to implement other tasks of interest with significantly less quantum resource. As examples, we show that the algorithm can be applied to reduce the data dimensionality of two well-known quantum machine learning algorithms, quantum support vector machine and quantum linear regression for prediction. This demonstrates that our algorithm has the potential to be well adapted to QML, and to free it from the curse of dimensionality to solve some problems
of practical importance. Since PCA underlies a variety of linear or nonlinear DR algorithms, our quantum algorithm may be promising to inspire more
quantum data compression algorithms based on other linear or nonlinear DR approaches.

\section*{Acknowledgements}
We thank Samuel Marsh, Amit Sett, Mitchell Chiew and Kooper De Lacy for helpful discussions. This work is supported by NSFC (Grant Nos. 61572081, 61672110, and 61671082). C.-H. Yu is supported by China Scholarship Council.


\end{document}